\documentclass[12pt]{amsart}

\usepackage{amsmath}
  \usepackage{paralist}
  \usepackage{graphics} 
  \usepackage{epsfig} 
 \usepackage[colorlinks=true]{hyperref}
\hypersetup{urlcolor=blue, citecolor=red}

\def\currentvolume{X}
 \def\currentissue{X}
  \def\currentyear{200X}
   \def\currentmonth{XX}
    \def\ppages{X--XX}

\newtheorem{theorem}{Theorem}[section]
\newtheorem{corollary}[theorem]{Corollary}

\newtheorem{lemma}[theorem]{Lemma}

\theoremstyle{definition}

\newtheorem{example}[theorem]{Example}

\title[Topological entropy of unimodal maps]
      {Computing the topological entropy of unimodal maps}

\author[Rui Dil\~{a}o and Jos\'{e} Amig\'{o}]{}


 \email{jm.amigo@umh.es}
 \email{rui@sd.ist.utl.pt}

\thanks{J. A. has been supported by the Spanish Ministry of Science and Innovation grant   MTM2009-11820.}

\begin{document}
\maketitle 

\centerline{\scshape Rui Dil\~{a}o}
\medskip
{\footnotesize
 \centerline{ NonLinear Dynamics Group, IST, Department of Physics}
   \centerline{Av. Rovisco Pais, 1049-001 Lisbon, Portugal}
} 

\medskip

\centerline{\scshape Jos\'{e} Amig\'{o}}
\medskip
{\footnotesize
 \centerline{Centro de Investigaci\'{o}n Operativa, Universidad Miguel Hern\'{a}ndez}
 \centerline{Av. de la Universidad s/n, 03202 Elche, Spain}
}

\bigskip

 \centerline{(Communicated by the associate editor name)}

\begin{abstract}
We derive an algorithm to determine recursively the lap number (minimal
number of monotone pieces) of the iterates of unimodal maps of an interval with free end-points. The algorithm is obtained
by the sign analysis of the itineraries of the critical point and of the boundary
points of the interval map. We apply this algorithm to the estimation of the growth number and the topological entropy of maps with direct and reverse bifurcations.
\end{abstract}

\section{Introduction} \label{sec1}

Topological entropy, introduced in 1965 by Adler, Konheim and McAndrew,
\cite{Adler}, is an invariant of topological conjugacy for self-maps of an
interval. Here, using symbolic dynamical
techniques, we derive a formula to calculate the lap number of the iterates of
the class of maps of an interval  with free end-points, leading to a straightforward estimation
of the topological entropy of these maps. The lap numbers of this class of maps depends on the symbolic itineraries of the critical point and of the two extreme points. 

Let $I=[a,b]$ be a closed interval of the real line $\mathbb{R}$. We
consider the class $\mathcal{F}$ of $C^{2}(I)$ maps $f:I\rightarrow I$, with a critical point at $x_{c}\in (a,b)$, and such that,

\begin{description}
\item[(I)] $f^{\prime \prime }(x_{c})<0$,

\item[(II)] $f^{\prime }(x)>0$ for $x<x_{c}$, and $f^{\prime }(x)<0$ for $%
x>x_{c}$.
\end{description}

In this paper, the single-humped maps in $\mathcal{F}$ are called \textit{%
unimodal maps}. The maps in the class $\mathcal{F}$ are not necessarily
symmetric around the critical point, and the iterates of the end-points of
the interval $I$ need not to converge to the same limit set.

The chain rule of differentiation applied to the $n$th iterate of $f$, written $%
f^{n}$ ($f^{0}$ is the identity map), shows trivially that along with $f$,
all iterates $f^{n}$ are also twice differentiable on $I$. In particular,
\begin{equation}
f^{n}{}^{\prime }(x)=f^{\prime }(f^{n-1}(x))f^{\prime }(f^{n-2}(x))\cdots
f^{\prime }(x)\,  \label{fmprima}
\end{equation}%
implies that $x_{c}$ is a critical point of $f^{n}$ for every $n\geq 1$.
From (\ref{fmprima}) and condition (II) we have:

\begin{lemma}
\label{lemma1} If $f\in \mathcal{F}$, then the critical points of $f^{n}$, $%
n> 1$, are the points $x\in (a,b)$ such that $f^{i}(x)=x_{c}$, for some $%
0< i\leq n-1$. Moreover, all the critical points of $f^{n}$, 
with $n\geq 1$, are maxima or minima, but not inflection points.
\end{lemma}

It follows from Lemma~\ref{lemma1}, that the critical points of $f^{n}$ with
$n\geq 1$, are the pre-images of the critical point $x_{c}$ up to order $n-1$%
. If $f(x_{c})>x_{c}$, $f(a)<x_{c}$ and $f(b)<x_{c}$, then the iterated maps
$f^{n}$, with $n\geq 2$, have more than one critical point. On the other
hand, if $f(x_{c})\leq x_{c}$, or $f(x_{c})>x_{c}$, $f(a)>x_{c}$ and $%
f(b)>x_{c}$, then the only critical point of $f^{n}$, with $n\geq 2$, is $%
x_{c}$.

A well-known example of a family of maps in $\mathcal{F}$ is provided by the
one-parameter quadratic (or logistic) family $f_{\mu }(x)=4\mu x(1-x)$,
where $I=[0,1]$ and the parameter $\mu \in (0,1]$. For this family of maps,
the critical point $x_{c}=1/2$ is $\mu $-independent. If $\mu \in (0,1/2]$,
then the only critical point of $f_{\mu }^n$, with $n\ge 1$, is $x=x_{c}=1/2$%
.

Two self-maps $f_{1}$ and $f_{2}$ of the intervals $I_{1}$ and $I_{2}$,
respectively, are topologically conjugate if there exists an homeomorphism $%
h:I_{1}\rightarrow I_{2}$ such that $f_{2}=h\circ f_{1}\circ h^{-1}$. The
topological entropy of a piecewise continuous interval map $f$ is an
invariant of topological conjugacy, \cite{Adler}, and its topological
entropy is calculated through the minimal number of monotone pieces or
laps of the iterates $f^{n}$, or by the number of fixed points of $f^{n}$,
\cite{Adler, Milnor,Misiu}. To be more precise, denoting by $\ell _{n}$ the
minimal number of monotone pieces of $f^{n}$, and by $N(f^{n})$ the
number of fixed points of $f^{n}$, then the topological entropy of $f$ is
given by, \cite{Misiu},
\begin{equation}
h(f)=\limsup_{n\rightarrow \infty }\frac{1}{n}\log \ell
_{n}=\limsup_{n\rightarrow \infty }\frac{1}{n}\log N(f^{n})  \label{topoent}
\end{equation}%
The number $s=\limsup_{n\rightarrow \infty } \ell_{n}^{1/n}$ is  called the growth number of $f$.
In the case of the logistic family of maps $f_{\mu }(x)=4\mu x(1-x)$, we have $h(f_{1})=\log 2$ 
  and $h(f_{\mu })=0$, for $\mu \in \lbrack 0,1/2]$.

An important characteristic of a unimodal map $f$ with positive topological
entropy is the existence of a semiconjugacy to the tent map with slopes $\pm
e^{h(f)}$. This is the Milnor and Thurston classification theorem, \cite%
{Milnor, Katok}. The variation of the topological entropy as a
parameter of a family of maps  is changed, implies that the number of periodic points of the
family of maps also changes, showing the existence of bifurcations in the
dynamics described by interval maps. This is in general associated with a
change in complexity in the sense of Li and Yorke's chaos scenario \cite%
{Li1975}. The  topological entropy for families of interval
maps is a way to analyze bifurcations  and is also a quantitative evaluation of the complexity of the dynamics.

Several authors have proposed approximating algorithms to calculate the
topological entropy of unimodal maps with fixed boundary points. In the case
of the logistic map, techniques based on Markov processes have been
introduced, \cite{Balm,Hof}. In this approach the spectral properties of a
transfer matrix with increasing dimension are analyzed. This transfer matrix
is defined by the images by the logistic map $f_{\mu }$, \cite{Balm,Hof}.
Another technique consists in approximating unimodal maps by piecewise
monotone maps and then calculating the topological entropy of the piecewise
approximation to the map, \cite{Block}. All these methods converge to the
topological entropy, however they are indirect and do not give  
information about the topological characteristics of the map, as, for
example, the lap numbers and the number of critical points.

A direct method of estimating and calculating the topological entropy of an interval map  is 
with the lap numbers $\ell _{n}$ of the
iterates of a map. In \cite{Dias1, Dias2} a recursive formula to calculate $%
\ell _{n}$ was proposed for the class of unimodal maps with fixed boundary
points. The proof for unimodal maps with a local minimum and fixed boundary
points was given  in \cite{Dilao}. Here, we generalize this
result for the larger class of maps $\mathcal{F}$. The main result of this
paper is Theorem~\ref{theorem3.4}, where we derive a general formula to
calculate recursively the lap number $\ell _{n}$ of the iterates of $f\in
\mathcal{F}$. This formula depends on the orbits (itineraries) of the critical point
and of the boundary points of the map. On the other hand, this formula can also be efficiently used to estimate the topological
entropy and the growth number for unimodal maps obtained by time series analysis.

The proof of Theorem~\ref{theorem3.4} is based on the
kneading calculus of Milnor-Thurston \cite{Milnor} and  on a symbolic
sequence introduced in \cite{Dias1} which is derived from the signed
itinerary of the critical point of a unimodal map. This new symbolic
sequence will be called the Min-Max sequence (MMS) of the map. The MMS gives
a procedure to determine on the interval $I$ the sequence of maxima and minima of any   
iterated map $f^n$, with $n\ge 1$.

To fix the notation, next, we describe  the symbolic approach due to Metropolis
\textit{et al}, \cite{Metropolis}, and recall some basics related to the
kneading calculus. The Min-Max sequence of a unimodal map and its most
important properties will be analyzed in the next section.

Consider a map $f\in \mathcal{F}$. Following Metropolis \textit{et al}, \cite%
{Metropolis}, we assign the symbols $I_{0}$ and $I_{1}$ to the intervals $%
[a,x_{c})$ and $(x_{c},b]$, respectively, and the symbol $C$ to the critical
point $x_{c}$. We  then associate to every $(f,x)\in \mathcal{F}\times I$%
, a sequence $\Theta _{f}(x)=(\Theta _{f,n}(x))_{n\geq 0}$ with entries $%
\Theta _{f,n}(x)$ belonging to the  
alphabet  $\ \mathcal{A}=\{I_{0},C,I_{1}\}$. The sequence $%
\Theta _{f}(x)$, called the \textit{itinerary} of $x$ under $f$, is defined
as follows:%
\begin{equation*}
\Theta _{f,n}(x)=\left\{
\begin{array}{ll}
I_{0} & \mbox{if }\text{ }f^{n}(x)<x_{c}, \\
C & \mbox{if }\text{ }f^{n}(x)=x_{c}, \\
I_{1} & \mbox{if }\text{ }f^{n}(x)>x_{c}.%
\end{array}%
\right.
\end{equation*}%
The space of all the itineraries will be denoted by $\mathcal{A}^{\mathbb{N}%
_{0}}$, where $\mathbb{N}_{0}=\{0,1,...\}$. Since,
\begin{equation}
\Theta _{f,n}(f(x))=\Theta _{f,n+1}(x),  \label{itf(x)}
\end{equation}%
the action of $f$ on $I$ translates into a left shift on $\mathcal{A}^{%
\mathbb{N}_{0}}$. The sequence $\gamma _{f}=(\gamma _{f,n})_{n\in \mathbb{N}%
} $, defined as,%
\begin{equation}
\gamma _{f,n}=\Theta _{f,n}(x_{c})=\Theta _{f,n-1}(f(x_{c})),\;\;n\geq 1,
\label{gammafn}
\end{equation}%
is called the \textit{kneading sequence} (KS for short) of $f$, \cite{Milnor}%
. Hence, $\gamma _{f}$ is the itinerary of the critical value $f(x_{c})$.

The \textit{signed itinerary} of $x\in I$ under the iteration of $f$, is the
sequence $\Theta _{f}^{\varepsilon }(x)=(\Theta _{f,n}^{\varepsilon
}(x))_{n\geq 0}$ on the extended alphabet $\mathcal{A}_{\varepsilon }\in
\{-I_{1},-C,-I_{0},0,I_{0},C,I_{1}\}$, where,
\begin{eqnarray} 
&\Theta _{f,0}^{\varepsilon }(x)=\Theta _{f,0}(x),\;\Theta
_{f,1}^{\varepsilon }(x)=\Theta _{f,1}(x),   \label{A}\\
&\Theta _{f,n}^{\varepsilon }(x)=\varepsilon (\Theta _{f,1}(x))...\varepsilon
(\Theta _{f,n-1}(x))\Theta _{f,n}(x)\;\;\mbox{for }\text{ }n\geq 2\text{,}
 \label{B}  
\end{eqnarray}%
and,
\begin{equation*}
\varepsilon (I_{0})=1,\varepsilon (C)=0,\text{\mbox{  and  }}\varepsilon
(I_{1})=-1.
\end{equation*}%
Note that, $\varepsilon (I_{0})$, $\varepsilon (C)$, and $\varepsilon (I_{1})$
coincide with the signs of $f^{\prime }$ on $I_{0}$, $C$, and $I_{1}$,
respectively.

We say that a sequence $\alpha \in \mathcal{A}_{\varepsilon }^{\mathbb{N}%
_{0}}$ is \textit{admissible} if there exists  $f\in \mathcal{F}$ such
that $\alpha =\Theta ^{\varepsilon }(f(x_{c}))$. From now on, we shall always
consider admissible sequences without explicitly stating it.

In the following, we write $\Theta (x)$ instead of $\Theta _{f}(x)$ whenever the
map $f$ is understood from the context, and the same applies to the kneading
sequence and signed itineraries.

\section{Geometry of the signed itineraries: The Min-Max sequence} \label{sec2}

We now relate the signed itineraries of the critical point of a unimodal map
$f\in \mathcal{F}$ with the structure of maxima and minima of the iterates of $%
f$.

Suppose that $f^{n}$, $n\geq 1$, has a local maximum (resp. minimum) at some
point $x\in I=[a,b]$. In order to simplify the notation and the proofs, in the
following, we say that $f^{n}(x)$ is a \textquotedblleft
positive\textquotedblright\ maximum (resp. minimum), if $f^{n}(x)-x_{c}>0$.
If, otherwise, $f^{n}(x)-x_{c}<0$, then we say that $f^{n}(x)$ is a
\textquotedblleft negative\textquotedblright\ maximum (resp. minimum). In
the remaining case, $f^{n}(x)-x_{c}=0$, and we say that $f^{n}(x)$ is a
\textquotedblleft zero\textquotedblright\ maximum (resp. minimum).

\begin{lemma}[\protect\cite{Dias1}]
\label{lemma2.1} Let $f\in \mathcal{F}$, and $k\geq 1$.  Then:

\begin{description}
\item[(a)] If $f^{k}(x)=x_{c}$ and $f(x_c)>x_c$, then $f^{k+1}(x)$ is a positive maximum.

\item[(b)] If $f^{k}(x)$ is a negative [resp. positive] minimum, then $%
f^{k+1}(x)$ is a minimum [resp. maximum].

\item[(c)] If $f^{k}(x)$ is a negative [resp. positive] maximum, then $%
f^{k+1}(x)$ is a maximum [resp. minimum].

\item[(d)] If  $f(x_c)\le x_c$, then $f^{k}(x)$ has only one critical point, a negative or a zero maximum.

\end{description}
\end{lemma}

\begin{proof} For $k\geq 1$ we have,
\begin{equation}
f^{k+1}{}^{\prime }(x)=(f\circ f^{k})^{\prime }(x)=f^{\prime
}(f^{k}(x))f^{k}{}^{\prime }(x)
\label{Df^(k+1)}
\end{equation}%
and,
\begin{equation}
f^{k+1}{}^{\prime \prime }(x)=f^{\prime \prime }(f^{k}(x))(f^{k}{}^{\prime
}(x))^{2}+f^{\prime }(f^{k}(x))f^{k}{}^{\prime \prime }(x).
\label{DF2^(k+1)}
\end{equation}

(a) If $f^{k}(x)=x_{c}$, by (II), $f^{k+1}{}^{\prime }(x)=f^{\prime
}(x_{c})f^{k}{}^{\prime }(x)=0$, and by (I), $f^{k+1}{}^{\prime \prime
}(x)=f^{\prime \prime }(x_{c})(f^{k}{}^{\prime }(x))^{2}<0$. As $f(x_c)>x_c$, $f^{k+1}(x)=f(x_{c})>x_{c}$.

(b) If $f^{k}(x)$ is a minimum, then $f^{k+1}{}^{\prime }(x)=f^{\prime
}(f^{k}(x))f^{k}{}^{\prime }(x)=0$, and,
\begin{equation*}
f^{k+1}{}^{\prime \prime }(x)=f^{\prime }(f^{k}(x))f^{k}{}^{\prime \prime
}(x),
\end{equation*}%
where $f^{k}{}^{\prime \prime }(x)>0$. Hence, $f^{k+1}{}^{\prime \prime }(x)$
has the same sign as $f^{\prime }(f^{k}(x))$. If $f^{k}(x)<x_{c}$, then, by
(II), $f^{k+1}{}^{\prime \prime }(x)>0$ and $f^{k+1}(x)$ is a minimum. Likewise,
if $f^{k}(x)>x_{c}$, then, by (II), $f^{k+1}{}^{\prime \prime }(x)<0$ and $f^{k+1}(x)$ is
a maximum.

The proof of (c) is similar and (d) follows from Lemma~\ref{lemma1}.
\end{proof}

Let $f\in \mathcal{F}$ and set,
\begin{equation}
\mathcal{S}^{k}=\{x\in I:x\mbox{ is a
critical point for }f^{k}\}.  \label{S^k}
\end{equation}%
In particular, $\mathcal{S}^{1}=\{x_{c}\}$. According to Lemma~\ref{lemma1}, for $k\ge 1$,
$\mathcal{S}^{k}$ contains $x_{c}$ and its preimages by the iteration of $f$
up to order $k-1$. This same lemma or Eq.~(\ref{Df^(k+1)}) imply that if $%
x\in (a,b)$ is a critical point of $f^{k}$, $k\geq 1$, then $x$ is also a
critical point of $f^{n}$ for $n\geq k$. Hence, $\mathcal{S}^{k}\subset
\mathcal{S}^{k+1}$. These critical points can be maxima or minima, and the
corresponding critical values can be greater than, equal to, or smaller than
the critical point $x_{c}$.

In order to distinguish all these possibilities, we introduce the new alphabet
\begin{equation}
\mathcal{M}=\{m^{-},m^{0},m^{+},M^{-},M^{0},M^{+}\}\,,  \label{alphaM}
\end{equation}%
where \textquotedblleft $m$\textquotedblright\ stands for minimum and
\textquotedblleft $M$\textquotedblright\ stands for maximum. The superscript
signs attached to $m$ and $M$ specify additionally whether the extreme in
question is positive, negative or zero in the sense explained above in the beginning of this section.

Now, we define the sequence $\omega _{f}=(\omega _{f,n})_{n\geq 1}\in \mathcal{%
M}^{\mathbb{N}}$ as follows:%
\begin{equation*}
\omega _{f,n}=\left\{
\begin{array}{ll}
m^{-} & \text{if }f^{n}(x_{c})\text{ is a ``negative" minimum},\text{ } \\
m^{+} & \text{if }f^{n}(x_{c})\text{ is a ``positive" minimum}, \\
m^{0} & \text{if }f^{n}(x_{c})\text{ is a ``zero" minimum}, \\
M^{0} & \text{if }f^{n}(x_{c})\text{ is a ``zero" maximum}, \\
M^{-} & \text{if }f^{n}(x_{c})\text{ is a ``negative" maximum}, \\
M^{+} & \text{if }f^{n}(x_{c})\text{ is a ``positive" maximum}.%
\end{array}%
\right.
\end{equation*}%
The sequence $\omega _{f}$ is called the \textit{Min-Max sequence} of $f\in
\mathcal{F}$, or MMS for short, \cite{Dias1, Dilao}. For example, if $f^{n}(x)$
is a positive maximum, then we say that the extremum $f^{n}(x)$ is of type $%
M^{+}$. As
usual, we say that $\omega _{i}$ and $\omega _{j}$ have opposite signs if
the product of their signs is strictly negative.



The geometric meaning of the MMS of $f$ is clear: The $n$th iterate of $f$, $%
f^{n}$, has a critical point of type $\omega _{n}$ $\in \mathcal{M}$ at $%
x=x_{c}$.

Furthermore, if $f\in \mathcal{F}$, according to (I),
\begin{equation}
\mbox{sign }f^{\prime \prime }(x_{c})=-1.  \label{sign0}
\end{equation}%
If $n\geq 2$ and $f^{k}(x_{c})\neq x_{c}$ for $1\leq k\leq n-1$, then, by (%
\ref{fmprima}) and (II) we have
\begin{equation*}
f^{n}{}^{\prime \prime }(x_{c})=f^{\prime }(f^{n-1}(x_{c}))\cdots f^{\prime
}(f(x_{c}))f^{\prime \prime }(x_{c})\neq 0,
\end{equation*}%
and hence
\begin{equation}
\mbox{sign }f^{n}{}^{\prime \prime }(x_{c})=-\varepsilon (\gamma
_{f,1})\cdots \varepsilon (\gamma _{f,n-1}),\;\;(n\geq 2).  \label{sign}
\end{equation}%
Comparison of (\ref{sign0})-(\ref{sign}) with (\ref{A})-(\ref{B})
leads to the conclusion that the symbols $\Theta _{f,n}^{\varepsilon
}(x_{c}) $ can be identified with the symbols $\omega _{f,n}$ as in Table~%
\ref{tab1}, provided $\Theta _{f,n}^{\varepsilon }(x_{c})\neq 0$, for $n\geq
1$. More specifically, by the rules of Lemma~\ref%
{lemma2.1},
the symbols $I_{0}$, $C$ and $I_{1}$ in $\Theta _{f,n}^{\varepsilon }(x_{c})$
translate into the superscript signs $-$, $0$ and $+$, in $\omega _{f,n}$,
respectively, while the signs $+$ and $-$ in $\Theta _{f,n}^{\varepsilon
}(x_{c})$ translate into the symbols $M$ and $m$ in $\omega _{f,n}$.
If, for some $k\geq 2$, $\gamma _{f,k}=C$ (thus $\varepsilon (\gamma
_{f,k})=0$ and $\Theta _{f,k+1}^{\varepsilon }(x_{c})=0$), then the
construction of the MMS for $n\geq k+1$ follows the rules of Lemma~\ref%
{lemma2.1}.
These rules are summarized in Table~\ref%
{tab2}.

\begin{table}[th]
\begin{center}
\begin{tabular}{|c|l|c|}
\hline
$\Theta _{f,n}^{\varepsilon }(x_{c})$ &  & $\omega_{f,n}$ \\ \hline\hline
$I_{1}$ & $\longleftrightarrow $ & $M^{+}$ \\ \hline
$-I_{1}$ & $\longleftrightarrow $ & $m^{+}$ \\ \hline
$I_{0}$ & $\longleftrightarrow $ & $M^{-}$ \\ \hline
$-I_{0}$ & $\longleftrightarrow $ & $m^{-}$ \\ \hline
$C$ & $\longleftrightarrow $ & $M^{0}$ \\ \hline
$-C$ & $\longleftrightarrow $ & $m^{0}$ \\ \hline
\end{tabular}%
\end{center}
\caption{Correspondence between the symbols of the signed itinerary of $%
x_{c} $ and the \textit{Min-Max sequence} (MMS).}
\label{tab1}
\end{table}


Moreover, Lemma~\ref{lemma2.1}(a-c) also implies that consecutive symbols in the
MMS obey the transition diagram in Table~\ref{tab2}.

\begin{table}[th]
\begin{center}
\begin{tabular}{|c|l|c|}
\hline
$\omega_{f,n}$ &  & $\omega_{f,n+1}$ \\ \hline\hline
$m^{0},M^{0}$ & $\longrightarrow $ & $M^{+}$ \\ \hline
$m^{+},M^{-}$ & $\longrightarrow $ & $M^{-,0,+}$ \\ \hline
$m^{-},M^{+}$ & $\longrightarrow $ & $m^{-,0,+}$ \\ \hline
\end{tabular}%
\end{center}
\caption{Possible transitions between consecutive symbols of the MMS.}
\label{tab2}
\end{table}

In the following, we use the shorthand notation $(\alpha _{0}...\alpha
_{n})^{\infty }$ for any symbolic sequence that repeats indefinitely the
block $(\alpha _{0}...\alpha _{n})$. A symbolic sequence of the form $%
(\alpha _{0}\ldots \alpha _{n}(\alpha_{n+1}\ldots \alpha_{n+k})^{\infty })$
is called eventually periodic.

\begin{lemma}
\label{lemma2.2} Let $\gamma $ and $\omega $ be the KS and MMS of $f\in
\mathcal{F}$, respectively. If $\gamma $ is periodic of period $k$, then $%
\omega $ is also periodic and has period $k$ or $2k$. If $\gamma $ is
eventually periodic, then $\omega $ is also eventually periodic.
\end{lemma}

\begin{proof}
Suppose that the KS $\gamma $ is periodic with
period $k$, $\gamma =(\gamma _{1}...\gamma _{k})^{\infty }$.
Without loss of generality assume that $f(x_c)>x_c$, otherwise all the KS have one of the forms $\gamma=(I_0)^{\infty }$ or $\gamma=(C)^{\infty }$.
Thus, $\gamma
_{k+1}=\gamma _{1}=I_{1}$, and $\Theta _{k+1}^{\varepsilon }(x_{c})=\pm
\gamma _{k+1}=\pm I_{1}$. From Table~\ref{tab1}, it follows that $\omega _{k+1}=M^{+}$
or $\omega _{k+1}=m^{+}$. Since $\omega _{1}=M^{+}$, in the first case, it follows that $\omega
$ has period $k$. In the second case,
since $\gamma _{k+j}=\gamma _{j}$ for $j\geq 1$, we deduce that $f^{\prime
}(f^{k+j}(x_{c}))$ and $f^{\prime }(f^{j}(x_{c}))$ have the same sign for $%
j=1,...,k$, hence, by (I), 
\begin{equation*}
f^{2k+1}{}^{\prime \prime }(x_{c})=f^{\prime }(f^{2k}(x_{c}))\cdots
f^{\prime }(f^{k+1}(x_{c}))f^{\prime }(f^{k}(x_{c}))\cdots f^{\prime
}(f(x_{c}))f^{\prime \prime }(x_{c})<0
\end{equation*}
and $f^{2k+1}(x_{c})$
is a maximum. By the periodicity of the KS, $\gamma
_{2k+1}=\gamma _{k+1}=I_{1}$, which means that $f^{2k+1}(x_{c})>x_{c}$, thus
$\omega _{2k+1}=M^{+}$.

The second part of the lemma follows readily.
\end{proof}

\begin{example}
\label{example2.3} We consider unimodal maps $f\in \mathcal{F}\cap C^{3}(I)$
with negative Schwarzian derivative,
\begin{equation*}
Sf(x)\equiv \frac{f^{\prime \prime \prime }(x)}{f^{\prime }(x)}-\frac{3}{2}%
\left( \frac{f^{\prime \prime }(x)}{f^{\prime }(x)}\right) ^{2}<0\, .
\end{equation*}%
Unimodal maps with negative Schwarzian derivative have some special
properties, like possessing at most one stable periodic orbit, \cite{Singer}.

It can be shown that
\begin{equation*}
(I_{1}I_{0}I_{1}I_{1})^{\infty },\;I_{1}I_{0}(I_{1})^{\infty
},\;(I_{1}I_{0}I_{1})^{\infty }\hbox{ and  }I_{1}(I_{0})^{\infty }
\end{equation*}%
are admissible KS for maps of this class, \cite{Guckenheimer}. Using the
rules (\ref{A})-(\ref{B}) to construct the respective signed KS
and the rules in Table~\ref{tab1}, the corresponding MMS are:
\begin{equation*}
\begin{array}{c}
(M^{+}m^{-}m^{+}M^{+}m^{+}M^{-}M^{+}m^{+})^{\infty
},\;M^{+}m^{-}(m^{+}M^{+})^{\infty },\;(M^{+}m^{-}m^{+})^{\infty }\; %
\hbox{ and  } \\
M^{+}(m^{-})^{\infty }.%
\end{array}%
\end{equation*}
\end{example}


Let $f\in \mathcal{F}$, $f:[a,b]\rightarrow \lbrack a,b]$, and let $\mathcal{%
S}^{k}$ be the set of critical points of $f^{k}$ (see (\ref{S^k})). Let $%
x_{k}$ [resp. $y_{k}$] be the leftmost [resp. rightmost] critical point of $%
f^{k}$, i.e.,
\begin{equation}
x_{k}=\min \mathcal{S}^{k},\;\;\;y_{k}=\max \mathcal{S}^{k},  \label{x_k}
\end{equation}%
for $k\geq 1$. The relation between the critical
points  of $f^{k}$ and $f^{k+1}$ in the interval $(x_{k},y_{k})$ is described by the following lemma.%

\begin{lemma}
\label{lemma2.4} Let $f\in \mathcal{F}$ and $z_{k,1}<z_{k,2}$ be two
consecutive critical points for $f^{k}$, with $k\geq 2$. Then,

\begin{description}
\item[(a)] If $f^{k}(z_{k,1})$ and $f^{k}(z_{k,2})$ have opposite signs,
then there exists one and only one $z_{k+1}\in (z_{k,1},z_{k,2})$ such that $%
f^{k+1}$ has a positive maximum at $z_{k+1}$. Furthermore, $%
f^{k+1}(z_{k+1})=f(x_{c})$.

\item[(b)] Otherwise, there is no critical point of $f^{k+1}$ on $%
(z_{k,1},z_{k,2})$.
\end{description}
\end{lemma}

\begin{proof} We may assume that $f(x_c)>x_c$ (otherwise, by Lemma~\ref{lemma2.1}d), $f^{k}$ has only one critical point).
(a) By assumption, $f^{k}$ is monotone on $%
[z_{k,1},z_{k,2}]$. Suppose $f^{k}(z_{k,1})<x_{c}<f^{k}(z_{k,2})$. By the Mean
Value Theorem, there exists one and only one $z\in (z_{k,1},z_{k,2})$ such
that $f^{k}(z)=x_{c}$. Set $z=z_{k+1}$. Hence, by  (I) and as $f(x_c)>x_c$,
$f^{k+1}(z)=f(x_{c})>x_{c}$,
\begin{equation*}
f^{k+1}{}^{\prime }(z)=f^{\prime }(f^{k}(z))f^{k}{}^{\prime }(z)=f^{\prime
}(x_{c})f^{k}{}^{\prime }(z)=0,
\end{equation*}%
and, $f^{k+1}{}^{\prime \prime }(z)=f^{\prime \prime
}(x_{c})(f^{k}{}^{\prime }(z))^{2}<0.$ Therefore, $f^{k+1}$ has a positive
maximum at $z_{k+1}$. The case,
$f^{k}(z_{k,1})>x_{c}>f^{k}(z_{k,2})$ is dealt similarly.

(b) This assertion is straightforward.
\end{proof}

\section{Counting laps} \label{sec3}

Before analyzing the general case $f\in \mathcal{F}$ without any particular
boundary conditions, we consider  provisionally the following additional
boundary condition:

\begin{description}
\item[(III)] $f^{k}(a)<x_{c}\text{ and }f^{k}(b)<x_{c}$, $\text{for every }%
k\geq 1$
\end{description}

If $f\in \mathcal{F}$ and $f(x_{c})\leq x_{c}$, then the boundary conditions
(III) are trivially fulfilled and, as mentioned in the Introduction, $x_{c}$
is the only critical point of $f^{k}$ for all $k\geq 1$. This means that $%
x_{k}=y_{k}=x_{c}$ for all $k\geq 1$, see (\ref{x_k}). 

Interesting dynamics is  possible if $f(x_{c})>x_{c}$. In
this case, the boundary conditions (III) imply that the orbits of the
endpoints of the interval $I$, $a$ and $b$, are confined within the interval $[a,x_{c})$. This
happens, in particular, for maps anchored at the boundary points of the interval $[a,b]$, 
with $f(a)=f(b)=a$, as is the case of the logistic map.

\begin{lemma}
\label{lemma3.1} Let $f\in \mathcal{F}$ with $f(x_{c})>x_{c}$, and suppose
further that condition (III) is fulfilled. Let $\mathcal{S}^{k}$ be the set
of critical points of $f^{k}$, and let $x_{k}$ and $y_{k}$, $k\geq 1$, be
the leftmost and rightmost critical points of $f^{k}$, respectively. Then,

\begin{description}
\item[(a)] $f^{k}$ has a positive maximum at $x_{k}$ with (i) $x_{k}<x_{k-1}$,
for $k\geq 2$, and (ii) $f^{k}(x_{k})=f(x_{c}).$

\item[(b)] $f^{k}$ has a positive maximum at $y_{k}$ with (i) $y_{k}>y_{k-1}$,
for $k\geq 2$, and (ii) $f^{k}(y_{k})=f(x_{c}).$
\end{description}
\end{lemma}

\begin{proof} As $f$ is unimodal, $x_{1}=y_{1}=x_{c}$.

(a) Let $k=2$. Then, by (I) and (II), $f^{2\prime \prime }(x_{c})=f^{\prime
}(f(x_{c}))f^{\prime \prime }(x_{c})>0$, so $f^{2}(x_{c})$ is a minimum. Furthermore, the equation $f^{2}{}^{\prime }(x)=f^{\prime }(f(x))f^{\prime
}(x)=0$ has one and only one solution on $(a,x_{c})=(a,x_{1})$. In fact, by
(I) and (II), $f^{\prime }(x)>0$ for $x\in (a,x_{c})$, and $f^{\prime
}(f(x))=0$ implies that $f(x)=x_{c}$. By (III) and the condition $f(x_c)>x_c$, we have $f(a)<x_{c}<f(x_{c})$, and the equation $f(x)=x_{c}$ has a solution on
$(a,x_{c})$. This solution is unique due to the monotonicity of $f$ on the
interval $(a,x_{c})$. Hence, the unique critical point of $f^{2}$ on $(a,x_{c})$ is $x_{2}$, and $x_{2}=\min \mathcal{S}^{2}$.

We claim now that $f^{2}(x_{2})$ is a positive maximum.
Indeed, as $f(x_c)>x_c$,
\begin{equation*}
f(x_{2})=x_{c}\;\Rightarrow \;f^{2}(x_{2})=f(x_{c})>x_{c}\, ,
\end{equation*}%
and, by (I),
\begin{equation*}
f^{2}{}^{\prime \prime }(x_{2})=f^{\prime \prime }(x_{c})(f^{\prime
}(x_{2}))^{2}<0.
\end{equation*}

We proceed now by induction, supposing that (a) is true for $k=2,3,...,n-1$.
As by hypothesis, $f^{n-1}(x_{n-1})$ is a positive maximum, then $%
f^{n-1}(x_{n-1})>x_{c}$ and $f^{n-1}{}^{\prime \prime }(x_{n-1})<0$. Hence,
\begin{equation*}
f^{n}{}^{\prime \prime }(x_{n-1})=f^{\prime
}(f^{n-1}(x_{n-1}))f^{n-1}{}^{\prime \prime }(x_{n-1})>0\,.
\end{equation*}%
So, it follows that $f^{n}(x_{n-1})$ is a minimum. Furthermore, the
equation $f^{n}{}^{\prime }(x)=f^{\prime }(f^{n-1}(x))f^{n-1\prime }(x)=0$
has one and only one solution on the interval $(a,x_{n-1})$. Indeed, (i) $%
f^{n-1\prime }(x)>0$ on $(a,x_{n-1})$ because $x_{n-1}$ is the leftmost
critical point of $f^{n-1}$, and $f^{n-1}(x_{n-1})$ is a maximum; (ii)
by (III), the induction hypothesis and the monotonicity of $f^{n-1}$ on $(a,x_{n-1})$,
$f^{n-1}(x)=x_{c}$ has a unique solution on $(a,x_{n-1})$. We  call $%
x_{n}$ the unique critical point of $f^{n}$ on $(a,x_{n-1})$. Finally, to
prove that $f^{n}(x_{n})$ is a positive maximum, by (I) and as $f(x_c)>x_c$,
\begin{equation*}
f^{n-1}(x_{n})=x_{c}\;\Rightarrow \;f^{n}(x_{n})=f(x_{c})>x_{c},
\end{equation*}%
and,%
\begin{equation*}
f^{n}{}^{\prime \prime }(x_{n})=f^{\prime \prime }(x_{c})(f^{n-1}{}^{\prime
}(x_{n}))^{2}<0.
\end{equation*}

The proof of (b) is similar. \end{proof}

Given the KS of a map $f\in \mathcal{F}$, it is possible to sketch
symbolically and qualitatively the graph of $f^{k}$, for any $k\geq 1$.
Suppose first that $f$ obeys the boundary condition (III). We proceed as
follows:

\begin{description}
\item[(A)] Fix $k\geq 1$ and from the KS of $f\in \mathcal{F}$, construct
the first $k$ terms of the MMS, $\omega =(\omega _{1}=M^{+},\omega
_{2},...,\omega _{k})$.

\item[(B)] Draw two perpendicular coordinate axes and divide the vertical
axis into $k$ rows, corresponding, top to bottom, to the iterates $f^{i}$, $%
1\leq i\leq k$, Table~\ref{tab3}. The horizontal axis represents the interval $[a,x_{c}]$.
Divide the horizontal axis into $k+1$ columns, corresponding, right to left,
to the   critical points $x_{c}=x_{1}>x_{2}>...>x_{k}$ of Lemma~\ref{lemma3.1} and to the left endpoint $a$. On the leftmost column, above $a$,
enter on each row $i$ the sign of $f^{i}(a)-x_{c}$, for $1\leq i\leq k$ (i.e., ``$+$"
if $f^{i}(a)>x_{c}$, ``$0$" if $f^{i}(a)=x_{c}$, and ``$-$" if $f^{i}(a)>x_{c}$).
On the rightmost (or ``first") column,
above $x_{c}=x_{1}=y_{1}$, enter on each row $i$ the element $\omega _{i}$ of the
MMS, for $i=1,...,k$. 
Due to the boundary conditions (III) as assumed in Lemma \ref{lemma3.1},
the leftmost column of Table \ref{tab3} contains only minus signs.  However, dropping boundary condition (III), the left most column entries can have all the three signs. This construction is shown in Table~\ref{tab3}.

\item[(C)] On the second column of Table~\ref{tab3}, above $x_{2}$, enter $\omega _{i-1}$ on row
$i$, for $2\leq i\leq k$. Continue filling out the remaining columns above 
$x_{j}$. In the step $j$, enter $\omega _{i-j+1}$ on row $i$, for $j\leq
i\leq k$. In this way, we get a triangular matrix, with the symbol $M^{+}$
along the secondary diagonal, and each column being the down shift by one entry
of the next column to the right. This construction  shown in Table~\ref{tab3}, 
is based on the geometrical meaning of the MMS and on 
Lemma~\ref{lemma3.1}(a). In view of Lemma \ref{lemma3.1}(b), we can extend the horizontal axis
to include the interval $(x_{c},b]$, but, due to the boundary condition (III), we do not get additional
information. Up to this point, we know the structure of the local extrema of
$f^{i}$ at the special critical points $x_{1},...,x_{k}$ (and $%
y_{1},...,y_{k}$).
\end{description}

\begin{table}[th]
\begin{center}
\begin{tabular}{c|c|c|c|c|c||c}
- &  &  &  &  & $\omega_{1}$ & $1$ \\ \hline
- &  &  &  & $\omega_{1}$ & $\omega_{2}$ & $2$ \\ \hline
- &  &  & $\omega_{1}$ & $\omega_{2}$ & $\omega_{3}$ & $3$ \\ \hline
- &  &  & $\vdots $ & $\vdots $ & $\vdots $ & $\vdots $ \\ \hline
- & $\omega_{1}$ & $\cdots $ & $\omega_{k-2}$ & $\omega_{k-1}$ & $\omega_{k}$
& $k$ \\ \hline\hline
\hbox{$a$} & $x_{k}$ & $\cdots $ & $x_{3}$ & $x_{2}$ & $x_c=x_{1}=y_{1}$ &
\end{tabular}%
\end{center}
\caption{Ordering of some of the critical points of the iterates $f^{i}$ on
the interval $[a,x_{c}]$. This ordering is derived from Lemma~\protect\ref%
{lemma3.1}, and the construction (A)-(C). On the leftmost
column, we show the signs of $f^{i}(a)-x_{c}$, determined in this case by
condition (III). Note that the iterates of the
leftmost and rightmost critical points of $f^{i}$ have the same signed
itinerary as the critical point of the unimodal map $f$.  However, the localization of the maxima and minima of the
iterated map $f^{i}$ is not complete. To finish it, we must incorporate the
results of Lemma~\protect\ref{lemma2.4}, as explained in step (D).}
\label{tab3}
\end{table}

\begin{description}
\item[(D)] In order to determine the location of the remaining critical
points of $f^{2}$,..., $f^{k}$ on the interval $(a,x_{c}$), we analyze  the symbols of the MMS in 
row 2 of Table~\ref{tab3}. Consider the (only) two consecutive elements of
the MMS on row 2, namely $\omega _{1}$ and $\omega _{2}$. If they have
opposite sign, then by Lemma~\ref{lemma2.4}(a), $f^{3}$ has a positive
maximum at $z_{3}\in (x_{2},x_{1})$. Thus, from row $3$ downwards, a new column  is inserted between
the $x_{2}$- and the $x_{1}$-columns. This new column contains the symbols, $%
\omega _{1},\omega _{2},...,\omega _{k-2}$ 
(see Example~\ref{example3.2} below). If the sign of the two consecutive
elements on this row are not opposite, we are in the conditions of Lemma~\ref%
{lemma2.4}(b) and no new column is created.
Next, go   to rows 3, 4, etc. and compare row-wise all pairs of
consecutive elements. If two consecutive elements on the same row have
opposite signs, then we proceed as before and, under application of Lemma~%
\ref{lemma2.4}(a), we insert a new column in-between, beginning on the next
row with the entry $\omega _{1}$, followed by $\omega _{2},\omega
_{3},...$ on the remaining entries of this new column. The resulting diagram
displays the qualitative structure of local maxima and minima of the
iterates $f^{i}$, with $1\leq i\leq k$. According to Lemma~\ref{lemma1}, all
critical points of $f^{i}$ are pre-images of $x_{c}$ up to order $i-1$.
\end{description}

We call the ``MM-table or the Min-Max table of $f$" a table like the one in Table~\ref{tab4},
extended to the whole interval $[a,b]$.

\begin{example}
\label{example3.2} Let $f\in \mathcal{F}$ obeying the boundary condition
(III), and suppose that the KS of $f$ is $\gamma =(I_{1}I_{1}C)^{\infty }$.
By the rules of Table~\ref{tab2}, the corresponding MMS is $\omega
=(M^{+}m^{-}m^{0})^{\infty }$. In order to know the qualitative shape of,
say, $f^{4}$, we follow the steps (A)-(D) above. The symbolic table of
maxima and minima is represented in Table~\ref{tab4}. Row 4 of Table~\ref%
{tab4} provides the necessary information to draw qualitatively the graph of
$f^{4}$. From this construction, we can calculate directly the lap numbers of
$f^{k}$ for $1\leq k\leq 4$. From Table~\ref{tab4}, and counting also laps on the
right hand side of the critical point $x_{c}$, we obtain, $\ell _{1}=2$, $%
\ell _{2}=4$, $\ell _{3}=8$ and $\ell _{4}=14$.

\begin{table}[th]
\begin{center}
\begin{tabular}{cccccccccccc||c}
- &  &  &  &  &  &  &  &  &  &  & $M^{+}$ & $1$ \\ \hline
- &  &  &  &  &  & $M^{+}$ &  &  &  &  & $m^{-}$ & $2$ \\ \hline
- &  &  & $M^{+}$ &  &  & $m^{-}$ &  &  & $M^{+}$ &  & $m^{0}$ & $3$ \\
\hline
- & $M^{+}$ &  & $m^{-}$ &  & $M^{+}$ & $m^{0}$ & $M^{+}$ &  & $m^{-}$ &  & $%
M^{+}$ & $4$ \\ \hline\hline
\hbox{$a$} & $x_{4}$ &  & $x_{3}$ &  &  & $x_{2}$ &  &  &  &  & $x_c=x_{1}=y_{1}$
&
\end{tabular}%
\end{center}
\caption{Structure of the maxima and minima on the interval $[a,x_{c}]$ of
the first four iterates of a unimodal map $f$ obeying the boundary condition
(III), with KS $\protect\gamma =(I_{1}I_{1}C)^{\infty }$ and MMS $\protect%
\omega =(M^{+}m^{-}m^{0})^{\infty }$. After having constructed the table for the
first row ($k=1$), if two consecutive symbols of the row
number 1 have opposite signs, counting with the boundary conditions, then there is a new column beginning in the second row with
the first symbol of the MMS sequence. From this symbolic representation we can count
the number of laps of each iterated map. Due to the symmetry of the boundary
condition (III), this diagrams extends symmetrically with respect to the
critical point $x_{c}=x_{1}=y_{1}$ to the interval $[x_{c},b]$, and the
distribution of maxima and minima is mirrored. }
\label{tab4}
\end{table}
\end{example}

The construction just done provides the tools in order to prove Theorem~\ref%
{theorem3.4} of the next section, where we drop the boundary condition (III). The
existence of a simple zero of the equation $f^{n}(x)-x_{c}=0$ on $%
(a,x_{n-1}) $ [resp. on $(y_{n-1},b)$] depends on the signs of $%
f^{n}(a)-x_{c}$ and $f^{n}(x_{n-1})-x_{c}$ [resp. $f^{n}(y_{n-1})-x_{c}$ and
$f^{n}(b)-x_{c}$]. If the signs of, say, $f^{n}(a)-x_{c}$ and $%
f^{n}(x_{n-1})-x_{c}$ are opposite, then,  as in the proof of Lemma~%
\ref{lemma3.1}, it follows that there exists a unique $x_{n}\in (a,x_{n-1})$ such that $%
f^{n}(x_{n})=x_{c}$. Due to monotonicity, $x_{n}$ is a simple zero of $%
f^{n}(x)-x_{c}$. Otherwise, there is no such solution and so $x_{n}=x_{n-1}$%
, since $x_{n-1}$ is the leftmost critical point of $f^{n}$. This translates
\textit{mutatis mutandis} to the half interval $(x_{c},b)$. In the
following, we consider the general case, without assuming the boundary
conditions (III).

\begin{example}
\label{example3.5} We consider the map $f_{\alpha ,\beta }(x)=e^{-\alpha
^{2}x^{2}}+\beta \in \mathcal{F}$, with $\alpha =2.8$ and $\beta =-0.1$,
which will be analyzed in detail in  section~\ref{secapp}. This map has a
critical point at $x=0$. We suppose further that $f_{\alpha ,\beta
}(x):[-(1+\beta ),(1+\beta )]\rightarrow \lbrack -(1+\beta ),(1+\beta )]$,
and $a=-b=-(1+\beta )$. The first elements of the MMS of $f_{\alpha ,\beta }$
are $(M^{+}m^{-}m^{+}M^{-}M^{+}\ldots )$. The signs of the iterates $%
f_{\alpha ,\beta }^{n}(a)$ and $f_{\alpha ,\beta }^{n}(b)$, with $n\geq 1$,
are $(-,+,-,+,-,\cdots )$. With the construction done above in the steps (A)-(D), 
the MM-table of
the first five iterates of $f_{\alpha ,\beta }$ is represented in Table~\ref%
{tab5}. The difference from this case to the one presented in Example~\ref%
{example3.2} is  on the boundary points. Observe that 
$x_{3}=x_{2}$, since $f_{\alpha ,\beta }^{2}$ has a positive
maximum at $x_{2}$ and $f_{\alpha ,\beta }^{2}(a)>x_{c}$; $x_{4}=x_{3}$ because $f_{\alpha ,\beta }^{3}$ has a negative minimum at $x_3$ and $f_{\alpha ,\beta }^{3}(a)<x_{c}$.
The equality $x_{5}=x_{4}$ follows from a similar argument.

\begin{table}[th]
\begin{center}
\begin{tabular}{cccccc||c}
- &  &  &  &  & $M^{+}$ & 1 \\ \hline
+ & $M^{+}$ &  &  &  & $m^{-}$ & $2$ \\ \hline
- & $m^{-}$ &  &  & $M^{+}$ & $m^{+}$ & $3$ \\ \hline
+ & $m^{+}$ & $M^{+}$ &  & $m^{-}$ & $M^{-}$ & $4$ \\ \hline
- & $M^{-}$ & $m^{-}$ & $M^{+}$ & $m^{+}$ & $M^{+}$ & $5$ \\ \hline\hline
\hbox{$a$} & $x_{5}=\ldots =x_{2}$ &  &  &  & $x_c=x_{1}=y_{1}$ &
\end{tabular}%
\end{center}
\caption{Structure of the maxima and minima on the interval $[a,x_{c}]$ of
the first five iterates of the unimodal map $f_{\protect\alpha ,\protect%
\beta }$ of Example~\protect\ref{example3.5}. The first elements of the MMS
are $(M^{+}m^{-}m^{\ast }M^{-}M^{+}\ldots )$. Due to symmetry of $f_{\protect%
\alpha ,\protect\beta }$ with respect to $x_{c}=0$, the distribution of
maxima and minima of $f^i_{\protect%
\alpha ,\protect\beta }$ on $[x_{c},b]=[0,0.9]$ is the mirrored image of the maxima
and minima of $f^i_{\protect%
\alpha ,\protect\beta }$ on the interval $[a,x_{c}]=[-0.9,0]$.}
\label{tab5}
\end{table}
\end{example}

\section{Main results}\label{secmr}

Consider a unimodal map $f\in \mathcal{F}$, let $\ell _{n}$ denote the
\textit{minimal number of laps of} $f^{n}$, and let $e_{n}$ be the \textit{number
of local extrema of} $f^{n}$, with $n\geq 1$. Since $f^{n}$ is continuous
and piecewise monotone, the laps are separated by critical points, and the
relation,
\begin{equation}
\ell _{n}=e_{n}+1  \label{ln0}
\end{equation}%
holds.

Furthermore, let $s_{n}$ stand for the \textit{number of interior simple
zeros of} $f^{n}(x)-x_{c}$, $n\geq 1$, i.e., solutions of $f^{n}(x)=x_{c}$,
with $x\in (a,b)$, $f^{i}(x)\neq x_{c}$ for $0\leq i\leq n-1$, and $%
f^{n\prime }(x)\not=0$. In particular, for unimodal maps $f\in \mathcal{F}$,
we have $\ell _{1}=2$, $e_{1}=1$ and $s_{1}\in \{0,1,2\}$. If in addition
the boundary condition (III) is fulfilled, then $s_{1}=2$.

\begin{lemma}[\protect\cite{Dias1}]
\label{lemma3.3} Let $f\in \mathcal{F}$. Let $e_n$ be the number of critical 
points of $f^n$, and let $s_n$ be the number of interior simple zeros of $%
f^n(x)-x_c$. Then, for $n\geq 2$,
\begin{equation}
e_{n}=e_{n-1}+s_{n-1}.  \label{cn}
\end{equation}
\end{lemma}

\begin{proof} Indeed, $e_{n}$ equals the number of sign changes
of $f^{n\prime }$. Then, Eq. (\ref{cn}) follows from the relation $f^{n}{}^{\prime }(x)=f^{\prime
}(f^{n-1}(x))f^{n-1}{}^{\prime }(x)$.
\end{proof}

From (\ref{ln0}) and (\ref{cn}), we have,
\begin{equation}
\ell _{n}-\ell _{n-1}=e_{n}-e_{n-1}=s_{n-1}  \label{ln2}
\end{equation}%
for $n\geq 2$. Moreover, induction on $n$ with (\ref{cn}) yields,
\begin{equation}
e_{n}=e_{1}+\sum\limits_{i=1}^{n-1}s_{i}=1+\sum\limits_{i=1}^{n-1}s_{i}.
\label{cn2}
\end{equation}

For each iterate of a unimodal map, maxima and minima alternate
along the $x$ direction. This is easily seen in Table~\ref{tab4}. In the
simplest case, the row $k$ of the MM-table of $f\in \mathcal{F}$ contains
only the symbols $M^{+}$ and $m^{-}$. Therefore, the laps between two
consecutive extrema on the graph of $f^{k}$ always cross the line $y=x_{c}$.
Whether this also happens in the intervals $(a,x_{k})$ and $(y_{k},b)$
depends, of course, on the signs of $f^{k}(a)-x_{c}$ and $f^{k}(x_{k})-x_{c}$
on the left side of the interval $I$, and of $f^{k}(b)-x_{c}$ and $f^{k}(y_{k})-x_{c}$ on the right 
side of $I$. All the four possibilities are encapsulated in the relation,%
\begin{equation}
s_{k}=e_{k}+1-\alpha _{k}-\beta _{k},  \label{00}
\end{equation}%
where,
\begin{equation}
\begin{array}{l}
\alpha _{k}=\left\{
\begin{array}{rl}
0 & \hbox{if}\ \ \text{sign}(f^{k}(a)-x_{c})\cdot \text{sign}%
(f^{k}(x_{k})-x_{c})<0 \\
1 & \hbox{if}\ \ \text{sign}(f^{k}(a)-x_{c})\cdot \text{sign}%
(f^{k}(x_{k})-x_{c})\geq 0%
\end{array}%
\right.  \\[20pt]
\beta _{k}=\left\{
\begin{array}{rl}
0 & \hbox{if}\ \ \text{sign}(f^{k}(b)-x_{c})\cdot \text{sign}%
(f^{k}(y_{k})-x_{c})<0 \\
1 & \hbox{if}\ \ \text{sign}(f^{k}(b)-x_{c})\cdot \text{sign}%
(f^{k}(y_{k})-x_{c})\geq 0\, .%
\end{array}%
\right.
\end{array}
\label{000}
\end{equation}%

In the construction of the MM-table of $f$ of Example~\ref{example3.5}, we have
set $x_{i}=x_{i-1}$ ($i\geq 2$) every time $f^{i-1}(a)-x_{c}$ and $%
f^{i-1}(x_{i-1})-x_{c}$ have no opposite signs (see Table 5). The first
time this occurs, the type of the leftmost extreme of $f^i$ is not $\omega _{1}$
 but $\omega _{2}$; if $f^{i}(a)-x_{c}$ and $%
f^{i}(x_{i})-x_{c}$ have not opposite signs again, the type of the leftmost extremum of $f^{i+1}$ is $%
\omega _{3}$. In general, if $f^{i-1}(a)-x_{c}$ and $%
f^{i-1}(x_{i-1})-x_{c}$ have not opposite signs $j$ times, for $1\leq i\leq
k-1$, then the leftmost extremum $f^{k}(x_{k})$ is of type $\omega _{1+j}$
(see Table 5). Therefore, sign$(f^{k}(x_{k})-x_{c})=\,$sign$(\omega _{1+j})$.
 A similar discussion holds for the rightmost extremum $f^{k}(y_{k})$.

In general, the row $k$ contains any symbol from the whole alphabet $%
\mathcal{M}=$ $\{m^{-},m^{0},m^{+},M^{-},M^{0},M^{+}\}$. In this case, if
some symbol $\omega _{i}$ on the row $k$ belongs to the set,
\begin{equation*}
\mathcal{M}_{b}=\{m^{+},M^{-},m^{0},M^{0}\}
\end{equation*}%
then, none of the two entries adjacent to $\omega _{i}$ (eventually including
$f^{k}(a)-x_{c}$ or $f^{k}(b)-x_{c}$) can have a sign opposite to the sign
of $\omega _{i}$. Every time this happens,
there are two laps, one on each side of the extremum correponding to the
symbol $\omega _{i}\in \mathcal{M}_{b}$, which do not cross the line $y=x_{c}
$. If all such  symbols belonging to $\mathcal{M}_{b}$ correspond
to extrema other than the leftmost and rightmost extremun (so these are of
types $M^{+}$ or $m^{-}$), one gets from (\ref{00}),
\begin{equation}
s_{k}=e_{k}+1-\alpha _{k}-\beta _{k}-2b_{k}  \label{sn}
\end{equation}%
where $b_{k}$ is the number of symbols from $\mathcal{M}_{b}$ on row $k$ of
the MM-table of $f$ (see rows $k=3,4$ in Table 4 for an example on the
half-interval $[a,x_{c}]$). If, otherwise, $f^{k}(x_{k})$ (resp. $%
f^{k}(y_{k})$) is of a type included in $\mathcal{M}_{b}$, then we have to
set $\alpha _{k}=0$ (resp. $\beta _{k}=0$) in (\ref{sn}) since the term $%
2b_{k}$ already accounts for the absence of a simple zero of $f^{k}(x)-x_{c}$
in the interval $(a,x_{k})$ (resp. $(y_{k},b)$).

From the above discussion we conclude that Eq. (\ref{sn}) holds without
restrictions, provided that,
\begin{equation}
\begin{array}{l}
\alpha _{k}=\left\{
\begin{array}{rl}
0 & \hbox{if}\ \ \text{sign}(f^{k}(a)-x_{c})\cdot \text{sign}(\omega
_{1+\lambda (k-1)})<0 \\
0 & \hbox{if}\ \ \text{sign}(f^{k}(a)-x_{c})\cdot \text{sign}(\omega
_{1+\lambda (k-1)})\geq 0\ \hbox{and}\ \omega _{1+\lambda (k-1)}\in \mathcal{%
M}_{b} \\
1 & \hbox{if}\ \ \text{sign}(f^{k}(a)-x_{c})\cdot \text{sign}(\omega
_{1+\lambda (k-1)})\geq 0\ \hbox{and}\ \omega _{1+\lambda (k-1)}\not\in
\mathcal{M}_{b}%
\end{array}%
\right. \\[20pt]
\beta _{k}=\left\{
\begin{array}{rl}
0 & \hbox{if}\ \ \text{sign}(f^{k}(b)-x_{c})\cdot \text{sign}(\omega
_{1+\rho (k-1)})<0 \\
0 & \hbox{if}\ \ \text{sign}(f^{k}(b)-x_{c})\cdot \text{sign}(\omega
_{1+\rho (k-1)})\geq 0\ \hbox{and}\ \omega _{1+\rho (k-1)}\in \mathcal{M}_{b}
\\
1 & \hbox{if}\ \ \text{sign}(f^{k}(b)-x_{c})\cdot \text{sign}(\omega
_{1+\rho (k-1)})\geq 0\ \hbox{and}\ \omega _{1+\rho (k-1)}\not\in \mathcal{M}%
_{b}%
\end{array}%
\right.%
\end{array}
\label{delta}
\end{equation}%
where sign$(\omega _{i})=-1,0,1$ depending on whether $\omega _{i}\in
\{m^{-},M^{-}\}$, $\{m^{0},M^{0}\}$, or $\{m^{+},M^{+}\}$, respectively. The
functions $\lambda (k)$ and $\rho (k)$ are recursively calculated as
follows: $\lambda (0)=\rho (0)=0$, and for $k\geq 1$,
\begin{equation}
\begin{array}{l}
\lambda (k)=\left\{
\begin{array}{cl}
0 & \hbox{if}\ \ \text{sign}(f^{k}(a)-x_{c})\cdot \text{sign}(\omega
_{1+\lambda (k-1)})<0 \\
\lambda (k-1)+1 & \hbox{if}\ \ \text{sign}(f^{k}(a)-x_{c})\cdot \text{sign}%
(\omega _{1+\lambda (k-1)})\geq 0%
\end{array}%
\right. \\[20pt]
\rho (k)=\left\{
\begin{array}{cl}
0 & \hbox{if}\ \ \text{sign}(f^{k}(b)-x_{c})\cdot \text{sign}(\omega
_{1+\rho (k-1)})<0 \\
\rho (k-1)+1 & \hbox{if}\ \ \text{sign}(f^{k}(b)-x_{c})\cdot \text{sign}%
(\omega _{1+\rho (k-1)})\geq 0\,.%
\end{array}%
\right.%
\end{array}
\label{counters}
\end{equation}

We can now derive the main result of this paper.

\begin{theorem}\label{theorem3.4} Let $\omega =(\omega _{k})_{k\geq 1}$ be the MMS of $f\in
\mathcal{F}$, and set,
\begin{equation*}
J_{n}=\{1\leq j\leq n:\omega _{j}\in \mathcal{M}_{b}\}
\end{equation*}%
where $\mathcal{M}_{b}=\{m^{+},M^{-},m^{0},M^{0}\}$.
Let $(\alpha _{k})_{k\geq 1}$ and  $(\beta _{k})_{k\geq 1}$ be the sequences of
integers calculated recursively from (\ref{delta}) and (\ref{counters}),
then
\begin{equation}
\ell _{n+1}=2\ell _{n}{-2\sum\limits_{j\in J_{n}}(\ell _{n+1-j}-\ell _{n-j})}%
-\alpha _{n}-\beta _{n},  \label{ln1}
\end{equation}%
where $\ell _{0}=1$ and $n\geq 1$.
\end{theorem}

\begin{proof} Let $N_{p}$ be the number of columns that begin at
row $p$ on the MM-table of $f$, and, as before, let $b_n$ be the number  of symbols from the set $\{m^{+},M^{-},m^{0},M^{0}\}$ on row $n$ of the MM-table of $f$. Then,
\begin{equation}
b_{n}=\sum\limits_{j\in J_{n}}N_{n-j+1}.  \label{bn}
\end{equation}%
By Lemma~\ref{lemma2.4}(a), $N_{p}=s_{p-1}$. Upon substitution of this equality into (%
\ref{bn}) and (\ref{sn}), we obtain,
\begin{equation*}
s_{n}=e_{n}+1-2\sum\limits_{j\in J_{n}}s_{n-j} -\alpha_{n}-\beta_{n}
\end{equation*}%
From (\ref{cn2}) it follows now that,
\begin{equation}
s_{n}=2+\sum\limits_{i=1}^{n-1}s_{i}-2\sum\limits_{j\in J_{n}}s_{n-j}-\alpha_{n}-\beta_{n}.
\label{sn3}
\end{equation}%
Finally, with the condition $\ell_1=2$ and substitution of (\ref{sn3}) into (\ref{ln2}), we obtain   (\ref{ln1}).
\end{proof}

For example, we can check the  formulas 
(\ref{sn})-(\ref{ln1}) with the geometric
information contained in Table \ref{tab5} (extended to the half-interval $%
[x_{c},b]$), Example \ref{example3.5}. Here, $\alpha _{k}=\beta _{k}$ because
the map $f_{2.8,-0.1}$ of Example \ref{example3.5} is symmetric around the
critical point $x_{c}=0$,    $J_{1}=J_{2}=\emptyset $, $%
J_{3}=\{3\}$ and $J_{4}=\{3,4\}$. So, by Theorem~\ref{theorem3.4}, we obtain,
\begin{equation*}
\begin{tabular}{c|ccccc}
k & $\alpha _{k}=\beta _{k}$ & $b_{k}$ & $e_{k}$&$s_{k}$ & $\ell _{k}$ \\
\hline
$1$  & $0$ & $0$ & $1$& $2$ & $2$ \\
$2$  & $1$ & $0$ & $3$& $2$ & $4$ \\
$3$  & $1$ & $1$ & $5$& $2$ & $6$ \\
$4$  & $0$ & $3$ & $7$& $2$ & $8$ \\
$5$  & $0$ & $4$ & $9$& $2$ & $10$%
\end{tabular}%
\end{equation*}
where $b_k$ is the number of symbols of the set $\mathcal{M}_{b}$ of the row number $k$.

In the case of maps with the boundary condition (III), then $\alpha
_{k}=\beta _{k}=0$ for every $k\geq 1$ (see the proof of Lemma \ref{lemma3.1}
and Table \ref{tab4}). So, from (\ref{ln1}), we conclude:

\begin{corollary}[\protect\cite{Dias1, Dilao}]
\label{corollary3.5} Let $\omega =(\omega _{k})_{k\geq 1}$ be the MMS of $%
f\in \mathcal{F}$ fulfilling the boundary condition (III), and set,
\begin{equation*}
J_{n}=\{1\leq j\leq n:\omega _{j}\in \mathcal{M}_{b}\}\text{.}
\end{equation*}%
where $\mathcal{M}_{b}=\{m^{+},M^{-},m^{0},M^{0}\}$.
Then,
\begin{equation}
\ell _{n+1}=2\ell _{n}-2\sum\limits_{j\in J_{n}}(\ell _{n+1-j}-\ell _{n-j}),
\label{ln3}
\end{equation}%
where $\ell _{0}=1$ and $n\geq 1$.
\end{corollary}

The theorem~\ref{theorem3.4} and the corollary~\ref{corollary3.5} provide fast algorithms to calculate
lap numbers, thus allowing to estimate the topological entropy of smooth
unimodal maps in a simple and efficient way. Note that these formulas
involve the MMS of the map and the itineraries  of the endpoints of the interval $I$.

From the technical point of view, the conclusions of theorem~\ref{theorem3.4} remain true if we 
change the smoth conditions (I) a (II), by continuity and monotinicity conditions on the left and right sides of the critical point. 

\section{\textbf{Computing the topological entropy}}\label{secapp}

Theorem~\ref{theorem3.4} and Corollary~\ref{corollary3.5} are the basic
results in order to estimate the topological entropy of interval maps.

The first example we consider here is the one parameter family of logistic
maps, $x_{n+1}=f_{\mu }(x_{n})=4\mu x_{n}(1-x_{n})$, defined on the interval
$I=[0,1]$ and with $\mu \in (0,1]$. Clearly, $f_{\mu }\in \mathcal{F}$, for
all the values of $\mu \in (0,1]$, and the critical point of the family is
$x_{c}=1/2$, independently of $\mu $.

\begin{table}[!ht]
\begin{center}
{\footnotesize
\begin{tabular}{|c|c|c|c|c|}
\hline
$n$ & $\mu=0.875$ & $\mu=0.9196433776$ & $\mu=0.957$ & $\mu=1.0$ \\
& $(I_{1}I_{0}I_{1}I_{1})^{\infty }$ & $I_{1}I_{0}(I_{1})^{\infty }$ & $%
(I_{1}I_{0}I_{1})^{\infty }$ & $I_{1}(I_{0})^{\infty }$ \\
& $(M^{+}m^{-}m^{+}M^{+}m^{+}M^{-}M^{+}m^{+})^{\infty }$ & $%
M^{+}m^{-}(m^{+}M^{+})^{\infty }$ & $(M^{+}m^{-}m^{+})^{\infty }$ & $%
M^{+}(m^{-})^{\infty }$ \\ \hline
$1$ & 2 & 2 & 2 & 2 \\ \hline
$2$ & 4 & 4 & 4 & 4 \\ \hline
$3$ & 8 & 8 & 8 & 8 \\ \hline
$4$ & 14 & 14 & 14 & 16 \\ \hline
$5$ & 24 & 24 & 24 & 32 \\ \hline
$6$ & 38 & 38 & 40 & 64 \\ \hline
$7$ & 58 & 60 & 66 & 128 \\ \hline
\end{tabular}%
}
\end{center}
\caption{Lap number for several parameter values of the logistic map $f_{%
\protect\mu}(x)=4 \protect\mu x (1-x)$. For each parameter value, we show
the corresponding KS and MMS.}
\label{tab6}
\end{table}

To calculate the lap numbers for several values of the parameter $\mu $, we
choose $\mu =0.875$, $\mu =0.9196433776$, $\mu =0.957$ and $\mu =1$. For
these parameter values, the logistic maps have the KS and the
MMS of Example~\ref{example2.3} and,  in Table~\ref{tab6}, we show the first lap numbers calculated from corollary~\ref{corollary3.5}.

To calculate the growth numbers and the topological entropies for this family of maps, we can use theorem~\ref{theorem3.4} and corollary~\ref{corollary3.5} to calculate numerically the lap numbers and then  estimating $\limsup \ell_n^{1/n}$. As the rate of convergence of $\limsup \ell_n^{1/n}$ to its limit $s$ is not known, in order to have an estimate of the numerical errors $|s-\ell_n^{1/n}|$, we use the following  lemma adapted from \cite{Dilao}.

\begin{lemma}[\protect\cite{Dilao}]
\label{lemma4.1} Let $\omega =(\omega _{k})_{k\geq 1}$ be the MMS of $%
f\in \mathcal{F}$ fulfilling the boundary condition (III). Suppose in addition that the MMS has least period $k> 1$, $\omega=(\omega_1\omega_2\ldots\omega_k)^{\infty}$, and $f^k(x_c)\not=x_c$. Then, the growth number ($s$) of $f$ is the largest real root of the polynomial,
$$
P(x)=x^{k-1}-\sum_{i=1}^{k-1}\varepsilon_i x^{k-1-i}
$$
where,
$$
\varepsilon_i =\left\{\begin{array}{rl}
1&\hbox{ if  } \varepsilon_i = m^{-}\hbox{ or }M^{+}\\
-1&\hbox{ if  } \varepsilon_i = m^{+}\hbox{ or }M^{-}\, .
\end{array}\right.
$$
\end{lemma}

For the proof, See the Appendix~\ref{appen}.

To test the convergence of $(\ell _{n})^{1/n}$ to the growth numbers of the maps $f_{\mu}$, we take the MMS in Table~\ref{tab6}. For the first and third MMS, we use lemma~\ref{lemma4.1}, and we obtain, $s=1$ and $s=(1+\sqrt{5})/2$, respectively. The growth numbers of the second and forth MMS listed in Table~\ref{tab6} are,  $s=\sqrt{2}$ and $s=2$, \cite{Dias1}.
In the four cases, the growth numbers and the topological entropies are,
\begin{equation}
\begin{array}{cclcl}
(I_{1}I_{0}I_{1}I_{1})^{\infty }&:&   s=1&,&h=0\\
I_{1}I_{0}(I_{1})^{\infty }&:&   s=\sqrt{2}&,& h=\frac{1}{2}\log 2\\
(I_{1}I_{0}I_{1})^{\infty }&:&   s=\frac{1+\sqrt{5}}{2}&,&h= \log \frac{1+\sqrt{5}}{2}\\
I_{1}(I_{0})^{\infty }&:&   s=2&,& h= \log 2\, .
\end{array}
\label{eq-sh2}
\end{equation}

\begin{figure}[ht]
\begin{centering}
 \includegraphics[width= 0.7\hsize]{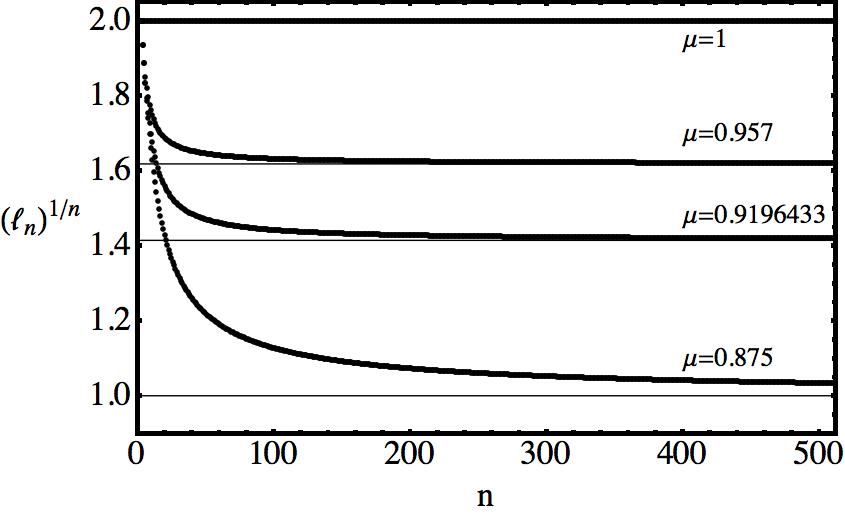}
 \caption{Behavior of the growth number $(\ell_n)^{(1/n)}$ as a function of $n$, for several parameter values of the logistic map $f_{\mu}(x )=4 \mu x  (1-x )$. The parameter values are the same as in Table~\ref{tab6}. The thin lines show the exact values of the growth numbers, calculated in (\ref{eq-sh2}). For $n=512$, the errors in the growth number estimates are below $\Delta s=0.005$, except for the first case of table Table~\ref{tab6}, where $\Delta s=0.034$.}
\label{fig1}
\end{centering}
\end{figure}

In Figure~\ref{fig1},  for the same parameter values
in Table~\ref{tab6}, we compare the behavior of  $(\ell _{n})^{(1/n)}$ as a function of $n$ with  the growth numbers (\ref{eq-sh2}). In Table~\ref{tab7}, we show the error between the exact and the numerical values of the growth number and of the topological entropy for the maps of Table~\ref{tab6}. In most cases, the error for the estimation of the topological entropy is acceptable for $n\simeq 512$. The worst estimate in Table~\ref{tab7}, correspond to the MMS of a dynamics on the Feigenbaum period doubling bifurcation cascade, prior to the appearance of fixed points with odd periods (\cite{Dias1}).

\begin{table}[!ht]
\begin{center}
\begin{tabular}{|c|c|c|c|c|}
\hline
 & $\Delta s=|s-\ell_n^{1/n}|$ & $\Delta h=|h-(\log\ell_n)/n|$  \\ \hline
$(I_{1}I_{0}I_{1}I_{1})^{\infty }$ & 0.034 & 0.033   \\ \hline
$I_{1}I_{0}(I_{1})^{\infty }$ & 0.005 & 0.004   \\ \hline
$(I_{1}I_{0}I_{1})^{\infty }$ & 0.003 & 0.002   \\ \hline
$I_{1}(I_{0})^{\infty }$ & 0 & 0   \\ \hline
\end{tabular}%
\end{center}
\caption{Error between the exact and the numerical values of the growth number and of the topological entropy for the maps of Table~\ref{tab6},  estimated with the choice $n=512$.}
\label{tab7}
\end{table}

\begin{figure}[ht]
\begin{centering}
 \includegraphics[width= 0.65\hsize]{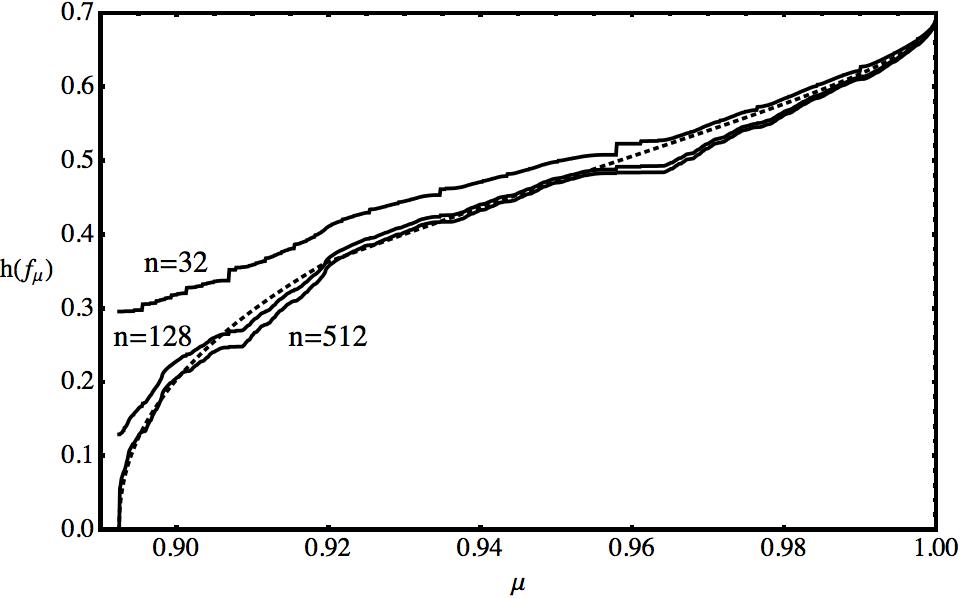}
 \caption{Topological entropy of the logistic map as a function of the parameter $\mu$, calculated from (\ref{topoent}), for $n=32$, $n=128$ and $n=512$. For $\mu \le \mu_{\infty }=0.892486416\ldots$, the topological entropy of the logistic family is zero, \cite{Dilao}. For $n=512$, we have imposed the condition $h(f_{\mu})=0$, for $\mu \le \mu_{\infty}$.
  The dotted line is  the  approximation (\ref{et}) to the topological entropy.}
\label{fig2}
\end{centering}
\end{figure}

In Figure~\ref{fig2}, we show the topological entropy of the logistic map as
a function of the parameter $\mu $ and calculated from (\ref{topoent}).  
We have calculated the KS and the MMS up to iterate number $n=512$,
and we have plotted $\log (\ell _{n})/n$ ($\simeq h$), for $n=32$, $n=128$ and $n=512$,
as a function of the parameter $\mu$. The lap number $\ell _{n}$ has been
calculated from Corollary~\ref{corollary3.5}. For $\mu \leq \mu _{\infty
}=0.892486416\ldots $ (the Feigenbaum point), the topological entropy of the
logistic map is zero (\cite{Dias2, Dilao}), and, for the case $n=512$, this
condition has been introduced by hand in the plot of Figure~\ref{fig2}. Near
$\mu_{\infty }$ the convergence of $\log (\ell _{n})/n$ to the topological
entropy is very slow, Table~\ref{tab7}. This can be seen in the approximations to the topological entropy with  $n=32$ and $%
n=128$. For these cases, we have not imposed the condition $h(f_{\mu})=0$, for $\mu \le \mu_{\infty}$. 

The monotonicity of the topological entropy for the logistic map, \cite{Douady} and \cite{Tsujii}, shows the
existence of bifurcations as the parameter $\mu $ is varied, and it measures
the complexity of the dynamics of the family of maps. In the chaotic region,
where the topological entropy is positive, these bifurcations are associated
with bifurcations of fixed points with large periods, which are  
difficult to detect numerically. The variation of the lap number and of the
topological entropy as the parameter $\mu $ changes, shows the existence of
bifurcations. In \cite{Dias2} and \cite{Dilao}, it has been shown that for
the logistic family of maps the topological entropy in the chaotic region is
well approximated by the function,
\begin{equation}
\tilde{h}(\mu )=\left\{
\begin{array}{ll}
& a\left( \mu -\mu _{\infty }\right) ^{b}\quad \hbox{if}\quad \mu \in
\lbrack \mu _{\infty },\eta _{2}] \\
& \log \left( 2-{\frac{2}{c(1-\mu )^{-d}-2}}\right) \quad \hbox{if}\quad \mu
\in \lbrack \eta _{2},1]%
\end{array}%
\right.  \label{et}
\end{equation}%
where $a=1.82957$, $b=\log 2/\log \delta =0.4498069$, $c=\pi /2=1.534780$, $%
d=1/2$, $\mu _{\infty }=0.892486416$, $\eta _{2}=0.919643377$ and $\delta
=4.6692\ldots $ is the Feigenbaum constant. In Figure~\ref{fig2}, we have
also plotted the approximation (\ref{et}) to the topological entropy.

\bigskip

We consider now the interval map $x_{n+1}=f_{\alpha ,\beta }(x_{n})$, where,
\begin{equation}
f_{\alpha ,\beta }(x)=e^{-\alpha ^{2}x^{2}}+\beta ,  \label{map2}
\end{equation}%
and $\alpha \neq 0$ and $\beta $ are real parameters. This two-parameter
family of interval maps qualitatively describes the dynamics of an
electrical circuit with a nonlinear diode showing chaotic behavior. The map (\ref{map2}) shows
direct and reverse bifurcations when the parameters are monotonically 
changed, \cite{Cascais}.

\begin{figure}[!ht]
\begin{centering}
 \includegraphics[width= 0.6\hsize]{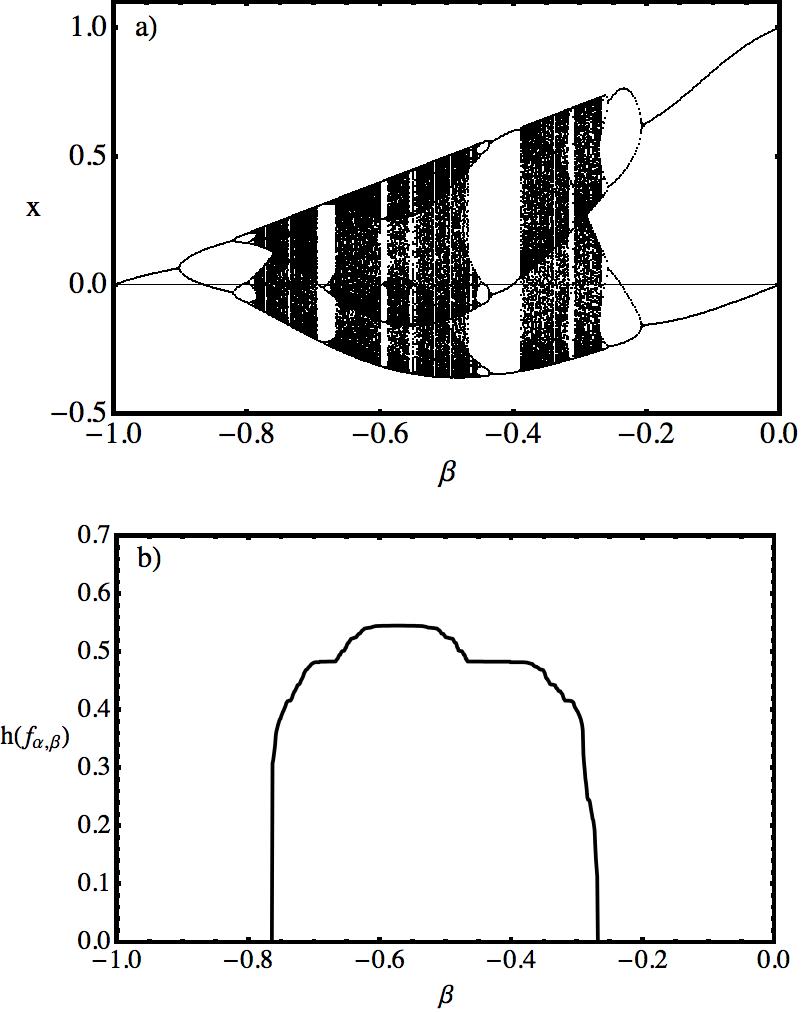}
 \caption{a) Bifurcation diagram of the map (\ref{map2}) as a function of the parameter $\beta$ in the interval $[-1,0]$ and for $\alpha=2.8$. b) Topological entropy of the map (\ref{map2}), with $\alpha=2.8$ and $\beta \in (-1,0]$. The topological entropy has been calculated by  (\ref{topoent}) with $n=512$, and the lap number has been calculated by Theorem~\ref{theorem3.4}. The parameter value where topological entropy changes from a  increasing to a decreasing function of $\beta$  corresponds to the reversion of bifurcations.}
\label{fig3}
\end{centering}
\end{figure}

The map (\ref{map2}) has a critical point at $x_{c}=0$. As $f_{\alpha ,\beta
}\in C^{\infty }(\mathbb{R})$, and $f_{\alpha ,\beta }^{\prime \prime
}(0)=-2\alpha ^{2}<0$, the map obeys condition (I). Condition (II) is
trivially fulfilled, and without loss of generality, we take $\alpha >0$. To
choose an interval $I$ such that $f_{\alpha ,\beta }$ maps $I$ to itself, we
take the non-trivial case $\beta \in(-1,0]$. As $f_{\alpha ,\beta
}(0)=1+\beta $, and,
\begin{equation*}
f_{\alpha ,\beta }^{2}(0)=\beta +e^{-\alpha ^{2}(1+\beta )^{2}}<1+\beta \ \
(>0)\, ,
\end{equation*}%
we can choose $I=[-(1+\beta ),(1+\beta )]$. Therefore, all the maps $%
f_{\alpha ,\beta }:[-(1+\beta ),(1+\beta )]\rightarrow \lbrack -(1+\beta
),(1+\beta )]$, with $\alpha >0$ and $\beta >-1$, are unimodal. As the
iterates of the boundary points $a=-(1+\beta )$ and $b=(1+\beta )$ can be on
both sides of the critical point $x_{c}=0$, we are in the conditions of
Theorem~\ref{theorem3.4}. Therefore, in order to estimate the topological
entropy of the map, we have to calculate numerically the iterates of the critical point and
the iterates of the boundary points of the interval $I$.

To better understand the relation between the topological entropy and the
dynamics generated by the family of maps (\ref{map2}), we have calculated an
estimate for the topological entropy through the exact value of the lap
number obtained from Theorem~\ref{theorem3.4}, and we have also computed the
bifurcation diagram of the dynamical system $x_{n+1}=f_{\alpha ,\beta
}(x_{n})$. In Figure~\ref{fig3}, we show the bifurcation diagram and the
topological entropy of the maps (\ref{map2}), for the parameter value $%
\alpha =2.8$ and $\beta\in [-1,0]$. As it is seen in Figure~\ref{fig3}, the region where the
topological entropy is increasing corresponds to an increase in the complexity
of the dynamics of the maps. This is due to the emergence of several periodic points.
When the reversal of bifurcations occur, the topological entropy decreases.

\section{\textbf{Conclusion}}

We have derived an algorithm to efficiently calculate the lap number and the topological entropy
of the class of single-humped unimodal maps with free end-points.  We have
shown that the lap number is determined by the itinerary of the critical
point and by the itinearies of the boundary points (Theorem \ref{theorem3.4}%
). The algorithm is based on
the geometric interpretation of the kneading sequence of a map, leading us
to a new symbolic sequence $(\omega _{f,n})_{n\in \mathbb{N}}$ --- the
Min-Max sequence --- that contains qualitative information about the
sequence of maxima and minima of the iterates  $f^{n}(x)$.

\appendix 
\section{Proof of Lemma~\ref{lemma4.1}}\label{appen}

\begin{proof}[Proof of Lemma~\ref{lemma4.1}]
As $f\in \mathcal{F}$,  fulfilling the boundary condition (III), we have $\alpha_k=\beta_k=0$,  and  by (\ref{sn3}), we obtain,
\begin{equation}
\begin{array}{lcl}
s_{n+k}-s_{n}&=&\sum\limits_{i=1}^{n+k-1}s_{i}-2\sum\limits_{j\in J_{n+k}}s_{n+k-j} -\sum\limits_{i=1}^{n-1}s_{i}+2\sum\limits_{j\in J_{n}}s_{n-j}\\
&=&\sum\limits_{i=n}^{n+k-1}s_{i}-2\sum\limits_{j\in J_{k}}s_{n+k-j}\, .
\end{array}
\label{app1}
\end{equation}%
As $\omega $ has least period $k$,  $\omega_1=\omega_{k+1}=M^{+}$ or $\omega_1=\omega_{k+1}=M^{-}$, and therefore, as in Table~\ref{tab2}, $\omega_{k}\in \mathcal{M}_{b}$. Introducing this condition in (\ref{app1}) and as $f^k(x_c)\not=x_c$, we get,
\begin{equation}
\begin{array}{lcl}
s_{n+k}&=&\sum\limits_{i=n+1}^{n+k-1}s_{i}-2\sum\limits_{j\in J_{k-1}}s_{n+k-j}\\
&=&\sum\limits_{j=1}^{k-1} \varepsilon_j s_{n+k-j}
\end{array}
\label{app2}
\end{equation}%
where,
$$
\varepsilon_i =\left\{\begin{array}{rl}
1&\hbox{ if  } \varepsilon_i = m^{-}\hbox{ or }M^{+}\\
-1&\hbox{ if  } \varepsilon_i = m^{+}\hbox{ or }M^{-}\, .
\end{array}\right.
$$
The recurrence relation (\ref{app2}), can be written in the matrix form,
\begin{equation}
\begin{pmatrix}s_{n+2}\\ \vdots \\ s_{n+k}\end{pmatrix}=
\begin{pmatrix}  0 & 1& 0& \cdots & 0\\
0 & 0& 1& \cdots & 0\\
\vdots & \vdots& \vdots& \cdots & 1\\
\varepsilon_{k-1} & \varepsilon_{k-2}& \varepsilon_{k-3} &\cdots &\varepsilon_{1}
\end{pmatrix}
\begin{pmatrix}s_{n+1}\\ \vdots \\ s_{n+k-1}\end{pmatrix} 
\label{app3}
\end{equation}
and $P(x)$ is the characteristic polynomial of the matrix in (\ref{app3}).
The solution of the recurrence relation (\ref{app3}) has the form,
\begin{equation}
s_n=\sum_{i=1}^{m} c_i p_i(n) \lambda_i^n
\label{app4}
\end{equation}
where, $\lambda_i$ are the eigenvalues of the characteristic polynomial $P(x)$, $c_i$ are constants and $p_i(n)$ are polynomials in $n$ of degree $m_i-1$, where $m_i$ is the multiplicity of $\lambda_i$. By (\ref{ln0}) and (\ref{cn}), we have,
\begin{equation}
\ell_n=2+\sum\limits_{i=1}^{n-1} s_{i}\, .
\label{app5}
\end{equation}
Substitution of (\ref{app4}) into (\ref{app5}), gives the result of the lemma.

\end{proof}

\end{document}